\documentclass[amsmath,amssymb,aps,prb,superscriptaddress,twocolumn]{revtex4-1}

\usepackage{hyperref}
\usepackage{latexsym,amsbsy,amsfonts,graphicx}
\usepackage{dcolumn}
\usepackage{bm}
\usepackage[utf8]{inputenc}
\usepackage[english]{babel}
\usepackage{color}
\usepackage{verbatim}

\usepackage{epstopdf}

\newcommand{\qv}{\mathbf{q}}
\newcommand{\kv}{\mathbf{k}}
\newcommand{\av}[1]{\ensuremath{\left\langle #1 \right\rangle}}

\begin{document}

\title{
From local to non-local correlations: the Dual Boson perspective
}

\author{E. A. Stepanov}
\affiliation{Radboud University, Institute for Molecules and Materials, 6525AJ Nijmegen, The Netherlands}

\author{A. Huber}
\affiliation{Institute of Theoretical Physics, University of Hamburg, 20355 Hamburg, Germany}

\author{E. G. C. P. van Loon}
\affiliation{Radboud University, Institute for Molecules and Materials, 6525AJ Nijmegen, The Netherlands}

\author{A. I. Lichtenstein}
\affiliation{Institute of Theoretical Physics, University of Hamburg, 20355 Hamburg, Germany}

\author{M. I. Katsnelson}
\affiliation{Radboud University, Institute for Molecules and Materials, 6525AJ Nijmegen, The Netherlands}

\date{\today}

\begin{abstract}
Extended dynamical mean-field theory (EDMFT) is insufficient to describe non-local effects in strongly correlated systems, since corrections to the mean-field solution are generally large.
We present an efficient scheme for the construction of diagrammatic extensions of EDMFT that avoids usual double-counting problem by using an exact change of variables (the Dual Boson formalism) to distinguish the correlations included in the dynamical mean-field solution and those beyond.
With a computational efficiency comparable to EDMFT$+$GW approach, our scheme significantly improves on the charge order transition phase boundary in the extended Hubbard model.
\end{abstract}

\maketitle

\section{Introduction}
The description of strongly correlated electronic systems is still one of the most challenging problems in condensed matter physics, despite a lot of efforts and plenty of suggested theories. One of the most popular approaches is the dynamical mean-field theory (DMFT)~\cite{PhysRevLett.62.324,RevModPhys.68.13,RevModPhys.78.865,WernerCasula}, which provides an approximate solution of the (multiband) Hubbard model by mapping it to a local impurity problem. Although DMFT neglects non-local correlation effects, it is able to capture important properties of the system such as the formation of Hubbard bands~\cite{Hubbard238,Hubbard401} and the Mott transition~\cite{mott1974metal,RevModPhys.70.1039}. Later, an extended dynamical mean-field theory (EDMFT)~\cite{PhysRevLett.77.3391, PhysRevB.61.5184, PhysRevLett.84.3678, PhysRevB.63.115110} was introduced to include collective (bosonic) degrees of freedom, such as charge or spin fluctuations, into DMFT. Unfortunately, these collective excitations have a strongly non-local nature, so a dynamical mean-field approach is insufficient and it was necessary to develop some extensions, we will call them EDMFT++, to treat non-local correlations.

The quantities of physical interest in EDMFT++ are the electronic self-energy $\Sigma_{\kv\nu}$ and polarization operator $\Pi_{\qv\omega}$.
The main idea of the dynamical mean-field approach is that all local correlations are already accounted for in the effective local impurity problem which results in the self-consistency conditions on the local part of lattice Green's function and susceptibility.
The mean-field ideology implies that in the EDMFT approach, the local self-energy and polarization operator are given by those of the impurity model.
To go beyond, one needs to determine the corrections $\bar{\Sigma}_{\kv\nu}$ and $\bar{\Pi}_{\qv\omega}$ to the electronic self-energy and polarization operator that describe non-local excitations.

However, as soon as one goes beyond the dynamical mean-field level, the \emph{non-local} corrections also change the \emph{local} parts of $\Sigma_{\kv\nu}$ and $\Pi_{\qv\omega}$. Indeed, the
self-consistency condition on the local part of the lattice
Green’s function $G_{\kv\nu}$ is not able to fix the local part of the self-energy $\Sigma_{\kv\nu}$ at the same time. Thus, the exact local part of full self-energy is no longer determined within the effective impurity problem and has contributions both from the dynamical mean-field solution and from the non-local corrections. The same holds true for polarization operator and the self-consistency condition on the local part of renormalized interaction. Therefore, great care should be taken to avoid double-counting of correlation effects when merging EDMFT with a diagrammatic approach.

The EDMFT$+$GW approach~\cite{PhysRevB.66.085120, PhysRevLett.90.086402, PhysRevLett.109.226401, PhysRevB.87.125149, PhysRevB.90.195114, PhysRevLett.113.266403, 2016arXiv160402023B} combines GW diagrams~\cite{PhysRev.139.A796, GW1, GW2} for the self-energy and polarization operator with EDMFT. In an attempt to avoid double-counting, all local contributions of the GW diagrams are subtracted and only the \emph{purely non-local} part of $\bar{\Sigma}_{\kv\nu}$ and $\bar{\Pi}_{\qv\omega}$ is used to describe non-local correlations. Exclusion of the impurity contributions from the diagrams introduced beyond EDMFT is necessary for a correct construction of the theory. However, the EDMFT$+$GW way of treating the double-counting problem is not unique and is the subject of hot discussions.

More complicated approaches invented to describe non-local effects with the impurity problem as a starting point are D$\Gamma$A~\cite{PhysRevB.75.045118}, 1PI~\cite{PhysRevB.88.115112} and DMF$^2$RG~\cite{PhysRevLett.112.196402}. These extensions of DMFT include two-particle vertex corrections in their diagrams. However, D$\Gamma$A and 1PI methods cannot describe the collective degrees of freedom arising from non-local interactions that are of interest here, and the DMF$^2$RG approach has not yet been applied to this problem. On the other hand, the recent TRILEX~\cite{PhysRevB.92.115109,2015arXiv151206719A} approach
was introduced to treat diagrammatically both fermionic and bosonic excitations. In this method the exact Hedin form~\cite{PhysRev.139.A796} of the lattice self-energy and polarization operator are approximated by including the full impurity fermion-boson vertex in the diagrams.

Instead of trying to construct the proper dynamical mean-field extension in terms of lattice Green's functions, one can take a different route and introduce so-called dual fermions (DF)~\cite{PhysRevB.77.033101} and dual bosons (DB)~\cite{Rubtsov20121320, PhysRevB.90.235135, PhysRevB.93.045107} and then deal with new \emph{dual} degrees of freedom. In these methods the local impurity model still serves as the starting point of a perturbation expansion, so (E)DMFT is reproduced as the non-interacting dual problem. It is important to point out that the self-energy and polarization operator in DF and DB are free from double-counting problems by construction: there is no overlap between the impurity contribution to the self-energy and polarization operator and local parts of dual diagrams since the impurity model deals with \emph{purely local} Green's functions only and the dual theory is constructed from \emph{purely non-local} building blocks. The impurity contribution has been excluded already on the level of the bare dual Green's function and interaction. Contrary to the existing methods, the DB approach does allow to describe strongly non-local collective excitations such as plasmons \cite{PhysRevLett.113.246407}.

The self-energy and polarization operator in self-consistent DB are built up as a ladder consisting of full fermion-fermion and fermion-boson vertices obtained from the local impurity problem. For computational applications, particularly those aimed at realistic multi-orbital systems, it can be convenient to use simpler approximations that do not require the computational complexity of the full two-particle vertex. To that end, we construct EDMFT++ schemes that do not require the full two-particle vertex, that exclude double-counting using the dual theory, and that contain the most essential parts of the non-local physics. We illustrate this by means of the charge-order transition in the extended Hubbard model.

\section{EDMFT++ theories}
The extended Hubbard model serves as the canonical example of a strongly correlated system where non-local effects play a crucial role. In momentum space, its action is given by the following relation
\begin{align}
S=-\sum_{{\bf k}\nu\sigma} c^{*}_{{\bf k}\nu\sigma}[i\nu+\mu-\varepsilon^{\phantom{*}}_{\bf k}]c^{\phantom{*}}_{{\bf k}\nu\sigma}+\frac{1}{2}\sum_{{\bf q}\omega}U^{\phantom{*}}_{\bf q}\rho^{*}_{\qv\omega}\rho^{\phantom{*}}_{\qv\omega}.
\label{eq:action}
\end{align}
Here we are interested only in the charge fluctuations, so in the following we suppress the spin labels on Grassmann variables $c^{*}_{\qv\nu}$ ($c^{\phantom{*}}_{\qv\nu}$) corresponding to creation (annihilation) of an electron with momentum $\kv$ and fermionic Matsubara frequency $\nu$. The interaction $U_{\bf q}=U+V_{\bf q}$ consists of the on-site $U$ and non-local interaction $V_{\qv}$ respectively. Here we consider $V_{\qv}$ as a nearest-neighbor interaction for simplicity. The charge fluctuations are given by the complex bosonic variable $\rho_{\omega}=~n_\omega-\av{n}\delta_{\omega}$, where $n^{\phantom{*}}_\omega = \sum_{\nu\sigma}c^{*}_{\nu}c^{\phantom{*}}_{\nu+\omega}$ counts the number of electrons and $\omega$ is a bosonic Matsubara frequency. The chemical potential $\mu$ is chosen in such a way that the average number of electrons per site is one (half-filling). Finally, $\varepsilon_{\kv}$ is the Fourier transform of the hopping integral $t$ between neighboring sites.

First of all, since we are interested in the EDMFT++ theories, let us briefly remind the main statements of the extended dynamical mean-field theory. In EDMFT, the lattice action \eqref{eq:action} is split up into a set of single-site local impurity actions $S_{\text{imp}}$ and a non-local remaining part $S_{\rm rem}$
\begin{align}
S =& \sum_{j} S^{(j)}_{\rm imp} + S^{\phantom{(j)}}_{\rm \hspace{-0.1cm}\phantom{i}rem},
\label{eq:actionlatt}
\end{align}
which are given by the following explicit relations
\begin{align}
S_{\text{imp}}=&-\sum_{\nu} c^{*}_{\nu}[i\nu+\mu-\Delta^{\phantom{*}}_{\nu}]c^{\phantom{*}}_{\nu} \notag\\
+&\, \frac{1}{2}\,\sum_{\omega}\,{\cal U}^{\phantom{*}}_{\omega}\, \rho^{*}_{\omega} \rho^{\phantom{*}}_{\omega}\label{eq:imp_action},\\
S_{\text{rem}} =&-\sum_{{\bf k}\nu} c^{*}_{{\bf k}\nu}[\Delta^{\phantom{*}}_{\nu}-\varepsilon^{\phantom{*}}_{\bf k}]c^{\phantom{*}}_{{\bf k}\nu} \notag\\
+&\, \frac{1}{2}\,\sum_{{\bf q}\omega}\,(U^{\phantom{*}}_{\bf q}-{\cal U}^{\phantom{*}}_{\omega})\, \rho^{*}_{{\bf q}\omega} \rho^{\phantom{*}}_{{\bf q}\omega}. \label{eq:rem_action}
\end{align}
Since the impurity model is solved exactly, our goal is to move most of the correlation effects into the impurity, so that the remainder is only weakly correlated. For this reason, two hybridization functions $\Delta_{\nu}$ and $\Lambda_{\omega}$ are introduced to describe the interplay between the impurity and external fermionic and bosonic baths respectively. These functions are determined self-consistently for an optimal description of local correlation effects. The local bare interaction of the impurity model is then equal to ${\cal U}_{\omega} = U + \Lambda_{\omega}$.
The impurity problem can be solved using, e.g., continuous-time quantum Monte Carlo solvers \cite{RevModPhys.83.349, PhysRevLett.104.146401}, and one can obtain the local impurity Green's function $g_{\nu}$, susceptibility $\chi_{\omega}$ and renormalized interaction ${\cal W}_{\omega}$ as
\begin{align}
g_{\nu} &= -\av{c^{\phantom{*}}_{\nu}c^{*}_{\nu}}_\text{imp}, \\
\chi_{\omega} &= -\av{\rho^{\phantom{*}}_{\omega}\rho^{*}_{\omega}}_\text{imp}, \\
{\cal W}_{\omega} &= {\cal U}_{\omega} + {\cal U}_{\omega}\chi_{\omega}{\cal U}_{\omega},
\end{align}
where the average is taken with respect to the impurity action \eqref{eq:imp_action}.
One can also introduce the local impurity self-energy $\Sigma_{\rm imp}$ and polarization operator $\Pi_{\rm imp}$
\begin{align}
\Sigma^{\phantom{*}}_{\rm imp} &= i\nu + \mu - \Delta^{\phantom{*}}_{\nu} - g^{-1}_{\nu}, \\
\Pi^{-1}_{\rm imp} &= \chi^{-1}_{\omega} + {\cal U}^{\phantom{*}}_{\omega},
\end{align}
that are used as the basis for the EDMFT Green's function $G_{\rm E}$ and renormalized interaction $W_{\rm E}$ defined as
\begin{align}
G_{\rm E}^{-1} &= \hspace{0.1cm}G_{0}^{-1}-\Sigma^{\phantom{1}}_{\rm imp}
= \hspace{0.1cm}g^{-1}_{\nu}-(\varepsilon^{\phantom{*}}_{\kv} - \Delta^{\phantom{*}}_{\nu}), \\
W_{\rm E}^{-1} &= W_{0}^{-1}-\Pi^{\phantom{1}}_{\rm imp}
= U_{\qv}^{-1}-(\chi^{-1}_{\omega}+{\cal U}^{\phantom{*}}_{\omega})^{-1}.
\end{align}
Here $G_{0}=(i\nu+\mu-\varepsilon_{\kv})^{-1}$ is the bare lattice Green's function and $W_{0}$ is the bare interaction, which is equal to $U_{\qv}$, or $V_{\qv}$ in the case of $UV$--, or $V$-- decoupling respectively \cite{PhysRevLett.109.226401, PhysRevB.87.125149}.

Importantly, a solution of every EDMFT++ theory can be exactly written in terms of EDMFT Green's functions and renormalized interactions as follows
\begin{align}
G_{\kv\nu}^{-1} &= \hspace{0.1cm}G_{0}^{-1}-\Sigma^{\phantom{1}}_{\kv\nu} \hspace{0.05cm}
= \hspace{0.1cm}G_{\rm E}^{-1}-\bar{\Sigma}^{\phantom{1}}_{\kv\nu},
\label{eq:nonlocal_Sigma} \\
W_{\qv\omega}^{-1} &= W_{0}^{-1}-\Pi^{\phantom{1}}_{\qv\omega}
= W_{\rm E}^{-1}-\bar{\Pi}^{\phantom{1}}_{\qv\omega},
\label{eq:nonlocal_Pi}
\end{align}
where $\Sigma_{\kv\nu}$ and $\Pi_{\qv\omega}$ are the exact, unknown in general, self-energy and polarization operator of the model respectively, and $\bar{\Sigma}_{\kv\nu}=~\Sigma_{\kv\nu}-~\Sigma_{\rm imp}$ and $\bar{\Pi}_{\qv\omega}=~\Pi_{\qv\omega}-\Pi_{\rm imp}$ are the corrections to the dynamical mean-field solution. With EDMFT as a starting point, the goal of EDMFT++ theories is to approximate these corrections. As pointed out above, $\bar{\Sigma}_{\kv\nu}$ and $\bar{\Pi}_{\qv\omega}$ should be introduced without double-counting with an effective local impurity problem, but still can give a local contributions to the lattice self-energy and polarization operator.

There is, in fact, a numerically exact way to obtain the non-local self-energy using the so-called bold diagrammatic Monte Carlo method~\cite{PhysRevB.83.161103}. However, this method is very expensive for realistic calculations, so we will be focused on less expensive diagrammatic methods.

\subsection{(E)DMFT$+$GW approach}
Historically, the EDMFT$+$GW approach~\cite{PhysRevB.66.085120, PhysRevLett.90.086402, PhysRevLett.109.226401, PhysRevB.87.125149, PhysRevB.90.195114} introduced the first approximations for $\bar{\Sigma}_{\kv\nu}$ and $\bar{\Pi}_{\qv\omega}$. Here, the self-energy and polarization operator diagrams from the GW approximation~\cite{PhysRev.139.A796, GW1, GW2} are added to the dynamical mean-field solution to treat non-local correlations,
\begin{align}
\Sigma^{\rm GW}_{\kv\nu} &=-\sum\limits_{\qv,\omega}G_{\kv+\qv,\nu+\omega}W_{\qv\omega},\label{eq:Sigma_Pi_GW}\\
\Pi^{\rm GW}_{\qv\omega} &= \,2\,\sum\limits_{\kv,\nu}\,G_{\kv+\qv,\nu+\omega}G_{\kv\nu},
\label{eq:Sigma_Pi_GW1}
\end{align}
where the coefficient ``2'' in Eq.~\eqref{eq:Sigma_Pi_GW1} accounts for the spin degeneracy. To avoid double-counting between the impurity correlations and the GW correlations, only the non-local part of Eqs.~\eqref{eq:Sigma_Pi_GW}-\eqref{eq:Sigma_Pi_GW1} is used, i.e., $\bar{\Sigma}^{\rm GW}_{\kv\nu}=~\Sigma^{\rm GW}_{\kv\nu} -~\Sigma^{\rm GW}_{\rm loc}$ and $\bar{\Pi}^{\rm GW}_{\qv\omega} = \Pi^{\rm GW}_{\qv\nu} - \Pi^{\rm GW}_{\rm loc}$. Since the local interaction $U$ has already been accounted for in the impurity problem, the bare non-local interaction in Eq.~\eqref{eq:Sigma_Pi_GW} can be taken in the form of $V$-- decoupling ($W_0 = V_{\qv}$), which leads to a simple separation of local and non-local contributions to the self-energy $\bar{\Sigma}_{\kv\nu}$. Unfortunately, this form of renormalized interaction overestimates non-local interactions~\cite{PhysRevLett.109.226401, PhysRevB.87.125149}. Alternatively, the form of $UV$-- decoupling ($W_0 = U_{\qv}$) is more consistent with standard perturbation theory for the full Coulomb interaction, but leads to the problems with separation of local and non-local parts of the diagrams. For example, it accounts only for the large local contribution ${\cal W}_{\omega}$ instead of the small full local four-point vertex function $\gamma^{4,0}$ as shown in Appendix~\ref{app:GW}. Therefore, the form of the renormalized interaction and the way to avoid the double-counting in general is a topic of hot discussions nowadays \cite{PhysRevB.91.235114}. 

Note that hereinafter the name $V$-- or $UV$-- decoupling in the EDMFT++ theories implies only the form of interaction $W_{0}$ used in the self-energy diagrams beyond the dynamical-field level. Since the aim of the paper is to compare the existing schemes of exclusion of the double-counting, the form of the self-energy diagrams in these both cases remains the same. Our notations can differ from those introduced in the previous works on EDMFT++ theories by the presence of additional diagrams in the different versions of decoupling schemes (see for example Ref.~\onlinecite{PhysRevB.87.125149}).

It should be noted, that there is another clear way to avoid the double-counting problem, namely simply ignoring non-local interactions in the dynamical mean-field part of the action and including them in the non-local corrections only. The impurity model then corresponds to DMFT, i.e., ${\cal U}_\omega = U$. Then, the non-local renormalized interaction in Eq.~\eqref{eq:Sigma_Pi_GW} be can taken in the form of $V$-- decoupling as $W_0 = V_{\qv}$, and the local part of this self-energy diagram is automatically zero. Although the DMFT$+$GW approach is free from double-counting by construction, it is less advanced than EDMFT+GW, since it ignores screening of the local interaction by non-local processes.

\subsection{Local vertex corrections beyond the EDMFT}
\label{sec:LocVertCor}
The exact self-energy and polarization operator of the lattice problem \eqref{eq:action} are given by the following relations~\cite{PhysRev.139.A796}
\begin{align}
\Sigma_{\kv\nu} &= -\sum\limits_{\qv\omega}G^{\phantom{\kv}}_{\kv+\qv,\nu+\omega}W^{\phantom{\kv}}_{\qv\omega}\Gamma^{\kv\qv}_{\nu\omega} =
\includegraphics[width=0.16\linewidth]{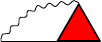}~, \label{eq:HedinSigma}\\
\Pi_{\qv\omega} &= \,2\,\sum\limits_{\kv\nu}G^{\phantom{\kv}}_{\kv+\qv,\nu+\omega}G^{\phantom{\kv}}_{\kv\nu}\,\Gamma^{\kv\qv}_{\nu\omega}=
\includegraphics[width=0.16\linewidth]{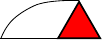}~, \label{eq:HedinPi}
\end{align}
where $\Gamma^{\kv\qv}_{\nu\omega}$ is the exact three-point Hedin vertex. Unfortunately, the full three-point vertex of the considered problem is unknown, and the self-energy and polarization operator can be found only approximately.
The most important correlation effects beyond EDMFT and the GW-diagrams are expected in the frequency-dependence of the fermion-boson vertex~\cite{PhysRevB.90.235135,PhysRevB.92.115109}. For this reason, the recent TRILEX~\cite{PhysRevB.92.115109,2015arXiv151206719A} approach with application to the Hubbard model was introduced. In this approach the exact Hedin vertex is approximated by the full local three-point vertex of impurity problem, which results in
\begin{align}
\Sigma^{\rm TRILEX}_{\kv\nu} &= -\sum\limits_{\qv\omega}G^{\phantom{\kv}}_{\kv+\qv,\nu+\omega}W^{\phantom{\kv}}_{\qv\omega}\gamma^{\phantom{1}}_{\nu\omega},\\
\Pi^{\rm TRILEX}_{\qv\omega} &= \,2\,\sum\limits_{\kv\nu}G^{\phantom{\kv}}_{\kv+\qv,\nu+\omega}G^{\phantom{\kv}}_{\kv\nu}\,\gamma^{\phantom{1}}_{\nu\omega},
\label{eq:TRILEX_diagr}
\end{align}
where $\gamma_{\nu\omega}$ is the full three-point vertex of the impurity problem determined below (see Eq.~\eqref{eq:3pvertex}). Thus, the local parts of the self-energy and polarization operator are identically equal to the local impurity quantities $\Sigma_{\rm imp}$ and $\Pi_{\rm imp}$ respectively. Moreover, it is possible to split $\Sigma^{\rm TRILEX}_{\kv\nu}$ and $\Pi^{\rm TRILEX}_{\qv\omega}$ into the local impurity part and non-local contribution as it was shown in Ref.~\onlinecite{2015arXiv151206719A}
\begin{align}
\Sigma^{\rm TRILEX}_{\kv\nu} &= \Sigma_{\rm imp} + \bar{\Sigma}^{\rm TRILEX}_{\kv\nu},\\
\Pi^{\rm TRILEX}_{\qv\omega} &= \Pi_{\rm imp} + \bar{\Pi}^{\rm TRILEX}_{\qv\omega},
\end{align}
where
\begin{align}
\bar{\Sigma}^{\rm TRILEX}_{\kv\nu} &= -\sum\limits_{\qv\omega}\bar{G}^{\rm TRILEX}_{\kv+\qv,\nu+\omega}\bar{W}^{\rm TRILEX}_{\qv\omega}\gamma^{\phantom{1}}_{\nu\omega},\\
\bar{\Pi}^{\rm TRILEX}_{\qv\omega} &= \,2\,\sum\limits_{\kv\nu}\bar{G}^{\rm TRILEX}_{\kv+\qv,\nu+\omega}\,\bar{G}^{\rm TRILEX}_{\kv\nu}\gamma^{\phantom{1}}_{\nu\omega},
\end{align}
and $\bar{G}^{\rm TRILEX}_{\kv\nu} = G_{\kv\nu}^{\phantom{1}}-g_{\nu}$, $\bar{W}^{\rm TRILEX}_{\qv\omega}=~W^{\phantom{1}}_{\qv\omega} - {\cal W}^{\phantom{1}}_{\omega}$ are non-local parts of the full lattice Green's function and renormalized interaction respectively.
Therefore, TRILEX approach is nothing more then an (E)DMFT+GW approximation with the same exclusion of double-counting, where the GW diagrams are additionally dressed with the local three-point vertex from one side. In this case, the lattice Green's function and renormalized interaction are given by the same Dyson Eqs.~\eqref{eq:nonlocal_Sigma}-\eqref{eq:nonlocal_Pi} with $\bar{\Sigma}^{\rm TRILEX}_{\kv\nu}$ and $\bar{\Pi}^{\rm TRILEX}_{\qv\omega}$ introduced beyond the dynamical mean-field level.

The main advantage of the TRILEX approach compared to existing diagrammatic methods is a computational efficiency due to the use of only the three-point vertex $\gamma_{\nu\omega}$ to threat non-local correlations. Nevertheless, even with this vertex function one can approximate the exact Hedin form of the self-energy and polarization function in a better way.
\begin{figure}[t!]
\flushleft{a) \,\,\,\, \includegraphics[width=0.85\linewidth]{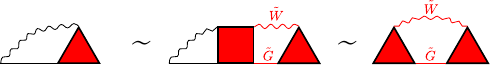}}~,
\flushleft{b) \,\,\,\, \includegraphics[width=0.85\linewidth]{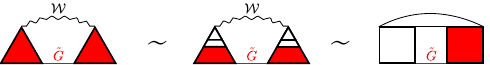}}~.
\caption{Hedin form of the self-energy diagram in case of a) at least one non-local Green's function $\tilde{G}$ and non-local renormalized interaction $\tilde{W}$, b) only local renormalized interactions ${\cal W}$. Straight and wave lines correspond to the Green's function and renormalized interaction.}
\label{fig:Hedin}
\end{figure}

It is of course true, that if the self-energy and polarization operator in the exact form of Eqs.~\eqref{eq:HedinSigma}-\eqref{eq:HedinPi} do not contain any non-local propagators, then these quantities are given by the impurity $\Sigma_{\rm imp}$ and $\Pi_{\rm imp}$ respectively. Therefore, the improvements concern only the contributions $\bar{\Sigma}^{\rm TRILEX}_{\kv\nu}$ and $\bar{\Pi}^{\rm TRILEX}_{\qv\omega}$, written in terms of non-local propagators and local impurity vertex functions.
As it was mentioned above, the self-consistency condition on the local parts of the Green's function and renormalized interaction cannot also fix the local parts of $\Sigma_{\kv\nu}$ and $\Pi_{\qv\omega}$ at the same time. Therefore, additional local contributions to the self-energy and polarization operator, hidden in the non-local structure of the exact three-point vertex, can appear from the diagrams introduced beyond the dynamical mean-field level. For example, the Hedin vertex with the same lattice indices at all three external points can contain non-local parts
\begin{align}
\includegraphics[width=0.55\linewidth]{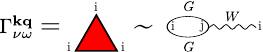}.
\label{eq:Vertex}
\end{align}
Therefore, these contributions are not provided by the local impurity problem and should be taken into account.

It is worth mentioning, that the Hedin form of the self-energy and polarization operator is exact for the theories with only one type of propagators. As soon as one includes the vertex functions of the impurity problem in the diagrams, all propagators become effectively separated into  two different types. Now, since the correction to the dynamical mean-field level is introduced in terms of only one (non-local) type of lines and all local lines are gathered in the local vertices, the Hedin form does not provide the exact result for the self-energy and polarization as shown in Refs.~\onlinecite{al1979contribution, AnKat}.

In order to discuss this in more detail, let us take a closer look at the Hedin diagram~\eqref{eq:HedinSigma} for the self-energy. Above we discussed the case of only local propagators. Now let us assume, that the Hedin vertex contains at least one non-local Green's function $\tilde{G}_{\kv\nu}$ and renormalized interaction $\tilde{W}_{\qv\omega}$. Then, the self-energy diagram can be reduced to the form of two renormalized three-point vertices with the non-local propagators in between as shown in Fig.~\ref{fig:Hedin} a). It may also  happen that one particular contribution to the lattice self-energy does not contain the non-local renormalized interaction at all. This case is shown in Fig.~\ref{fig:Hedin} b). The last case without a non-local Green's functions is not considered here due to appearance of higher-order vertex functions of the impurity problem in the diagrams.
The same procedure can be used for the polarization operator. Then, if we restrict ourselves only to the lowest order vertex function $\gamma_{\nu\omega}$, the self-energy and polarization operator introduced beyond the dynamical mean-field level are
\begin{align}
\includegraphics[width=0.3\linewidth]{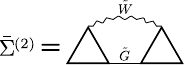}~,\label{eq:Sigma2}\\
\includegraphics[width=0.3\linewidth]{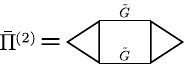}~,\label{eq:Pi2}
\end{align}
where, according to the above discussions, the three-point vertices appear at both sides of the GW diagrams, as was already discussed in Ref.~\cite{PhysRevB.66.085120}. Moreover, the specified form of the diagrams for the self-energy and polarization operator allows one to resum more
diagrams than with the use of TRILEX form.
\begin{figure}[t!]
\flushleft{a) \,\,\,\, \includegraphics[width=0.85\linewidth]{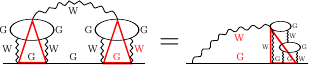}}~,\\
\vspace{-0.2cm}
\flushleft{b) \,\,\,\, \includegraphics[width=0.85\linewidth]{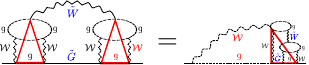}}~.
\caption{Structure of the vertex corrections in theories consisted of a) one and b) two types of propagators. Solid straight and wave lines correspond to the Green's function and renormalized interaction of one type and the dashed lines to those of the second type respectively.}
\label{fig:Two}
\end{figure}

The illustration of the importance of having the three-point vertex functions on both sides is also shown in Fig.~\ref{fig:Two}. The top row corresponds to a theory constructed from only one type of Green's functions. Then, the fermion-boson vertices are composed of the same propagators as the remainder of the diagram, and it is always possible to ``move'' all vertex correction to the right side of diagram and obtain the Hedin form for the self-energy~\cite{PhysRev.139.A796}. On the other hand, if the vertex functions are constructed from propagators (for example $g_{\nu}$ and ${\cal W}_{\omega}$ obtained from impurity problem) that differ from the Green's function $G$ and renormalized interaction $W$, it is no longer possible to obtain the Hedin form for this diagram. More clearly, the Hedin form is hidden inside of the impurity vertices. ``Moving'' the left part of the diagram to the right, as in the bottom row of Fig.~\ref{fig:Two}, gives a diagram with the same Hedin structure, but with different propagators.

So, if one prefers to work with the bare lattice propagators and to use the Hedin form of self-energy, then it would be consistent to approximate the exact Hedin vertex using the same bare lattice quantities without inclusion of any other types of propagators. If, instead, a combination of Green's functions and impurity vertices coming from different models is used, the renormalized vertices should be included at  both ends of the GW diagram for the self-energy and polarization operator.

In order to take the above corrections into account and to compare the double-counting exclusion schemes, one can introduce the EDMFT$+$GW$\gamma$ approach in the same way as EDMFT+GW by including the local impurity vertex $\gamma_{\nu\omega}$ in the GW diagrams as
\begin{align}
\Sigma^{{\rm GW}\gamma}_{\kv\nu} &= -\sum\limits_{\qv,\omega}\gamma_{\nu\omega}G_{\kv+\qv,\nu+\omega}W_{\qv\omega}\gamma_{\nu+\omega,-\omega},
\label{eq:Sigma_Pi_GWd}\\
\Pi^{{\rm GW}\gamma}_{\qv\omega} &= \,2\,\sum\limits_{\kv,\nu}\gamma_{\nu\omega}G_{\kv+\qv,\nu+\omega}\,G_{\kv\nu}\gamma_{\nu+\omega,-\omega}.
\label{eq:Sigma_Pi_GWd1}
\end{align}
Similarly to the EDMFT+GW case, only the non-local parts $\bar{\Sigma}^{{\rm GW}\gamma}_{\kv\nu}$ and $\bar{\Pi}^{{\rm GW}\gamma}_{\qv\omega}$ of the self-energy~\eqref{eq:Sigma_Pi_GWd} and the polarization operator~\eqref{eq:Sigma_Pi_GWd1} are used beyond the EDMFT. Then, the lattice quantities are given by the same equations~\eqref{eq:nonlocal_Sigma}-\eqref{eq:nonlocal_Pi}.

\section{Dual Boson approach}
A different way of accounting for non-local correlations beyond EDMFT is given by the Dual Boson approach~\cite{Rubtsov20121320,PhysRevB.93.045107}.
The non-local part $S_{\rm rem}$ of the lattice action \eqref{eq:action} can be rewritten in terms of new \emph{dual} variables $f^{*},f,\phi$ by performing a Hubbard--Stratonovich transformation, which leads to the dual action
\begin{align}
\tilde{S} &=
- \sum_{\kv\nu}f^{*}_{\kv\nu} \tilde{G}_{0}^{-1} f^{\phantom{*}}_{\kv\nu} - \frac{1}{2}\sum_{\qv\omega} \phi^{*}_{\qv\omega}\tilde{W}_{0}^{-1}\phi^{\phantom{*}}_{\qv\omega}+ \tilde{V}
\label{eq:dual_action}
\end{align}
with the bare dual propagators
\begin{align}
\tilde{G}_{0} &= \,G_{\rm E} - g_{\nu},
\label{eq:barefermionpropagator}\\
\tilde{W}_{0} &= W_{\rm E}-{\cal W}_{\omega},
\label{eq:barebosonpropagator}
\end{align}
and the full dual interaction $\tilde{V}$ that includes the impurity vertex functions $\gamma^{n,m}$ with $n$ fermion and $m$ boson lines to all orders in $n$ and $m$, as detailed in Appendix~\ref{ap:dualtr}. The first two terms in $\tilde{V}$ are given by the following relation
\begin{align}
\tilde{V}
&=\gamma^{2,1}_{\nu,\omega}\,f^{*}_{\nu}f^{\phantom{*}}_{\nu+\omega}\phi^{*}_{\omega}
 + \frac14\,\gamma^{4,0}_{\nu,\nu',\omega}\,f^{*}_{\nu}f^{*}_{\nu'}f^{\phantom{*}}_{\nu+\omega}
f^{\phantom{*}}_{\nu'-\omega}.
\end{align}
We define the three-point vertex $\gamma^{2,1}_{\nu\omega}$ in the same way as it is done in the TRILEX~\cite{PhysRevB.92.115109,2015arXiv151206719A} approach:
\begin{align}
\gamma^{2,1}_{\nu\omega}&=
g^{-1}_{\nu}g^{-1}_{\nu+\omega}\alpha^{-1}_{\omega}
\av{c^{\phantom{*}}_{\nu}c^{*}_{\nu+\omega}\rho^{\phantom{*}}_{\omega}},
\label{eq:3pvertex}
\end{align}
where $\alpha_{\omega}={\cal W}_{\omega}/{\cal U}_{\omega}=(1+{\cal U}_{\omega}\chi_{\omega})$ is the local renormalization factor. It is important to realize that this factor only affects the transformations from lattice to dual quantities and vice verse. Therefore, it does not change the final results written in terms of the initial lattice degrees of freedom. In order to shorten notations, hereinafter we call the three-point vertex as $\gamma_{\nu\omega}$. The four-point vertex function $\gamma^{4,0}_{\nu\nu'\omega}$ can be determined similarly to the previous papers on the Dual Boson formalism \cite{Rubtsov20121320,PhysRevB.93.045107}
\begin{align}
\gamma^{4,0}_{\nu\nu'\omega} = g^{-1}_{\nu}g^{-1}_{\nu'}g^{-1}_{\nu'-\omega}g^{-1}_{\nu+\omega} \Big[&\av{c^{\phantom{*}}_{\nu}c^{\phantom{*}}_{\nu'}c^{*}_{\nu'-\omega}
c^{*}_{\nu+\omega}}- \notag\\
&\,\,g_{\nu}g_{\nu'}(\delta_{\omega}-\delta_{\nu',\nu+\omega})\Big].
\end{align}
Then, the dual Green's function $\tilde{G}_{\kv\nu} =~ -\av{f^{\phantom{*}}_{\kv\nu}f^{*}_{\kv\nu}}$ and renormalized dual interaction $\tilde{W}_{\qv\omega} =~ -\av{\phi^{\phantom{*}}_{\qv\omega}\phi^{*}_{\qv\omega}}$, as well as dual self-energy $\tilde{\Sigma}_{\kv\nu}$ and polarization operator $\tilde{\Pi}_{\qv\omega}$, can be obtained diagrammatically~\cite{Rubtsov20121320, PhysRevB.90.235135, PhysRevB.93.045107}. These dual quantities have usual connection
\begin{align}
\tilde{G}_{\kv\nu}^{-1} &= \,\tilde{G}^{-1}_{0}\, - \tilde{\Sigma}^{\phantom{1}}_{\kv\nu},\\
\tilde{W}_{\qv\omega}^{-1} &= \tilde{W}^{-1}_{0} - \tilde{\Pi}^{\phantom{1}}_{\qv\omega}.
\label{eq:X-Pi}
\end{align}

To close the circle, the Green's function $G_{\kv\nu}$ and renormalized interaction $W_{\qv\omega}$ of the original model can be exactly expressed in terms of dual quantities via the similar Dyson Eqs.~\eqref{eq:nonlocal_Sigma}-\eqref{eq:nonlocal_Pi} as follows
\begin{align}
G_{\kv\nu}^{-1} &= \,G^{-1}_{\rm E}\, - \bar{\Sigma}^{\phantom{1}}_{\kv\nu},\label{eq:dual_G-Sigma}\\
W_{\qv\omega}^{-1} &= W^{-1}_{\rm E} - \bar{\Pi}^{\phantom{1}}_{\qv\omega},
\label{eq:dual_X-Pi}
\end{align}
where the self-energy and polarization operator introduced beyond EDMFT are
\begin{align}
\bar\Sigma^{\phantom{1}}_{\kv\nu} &= \frac{\tilde\Sigma_{\kv\nu}}{1+g_{\nu}\tilde\Sigma_{\kv\nu}},\label{eq:Sigmap}\\
\bar\Pi^{\phantom{1}}_{\qv\omega} &= \frac{\tilde\Pi_{\qv\omega}}{1+{\cal W}_{\omega}\tilde\Pi_{\qv\omega}}\label{eq:Pip}.
\end{align}

It should be noted, that the bare dual Green's function~\eqref{eq:barefermionpropagator} and renormalized interaction~\eqref{eq:barebosonpropagator} are strongly non-local due to the EDMFT self-consistency conditions
\begin{align}
\sum_{\kv}G_{\rm E}\, &= g_{\nu},\\
\sum_{\qv}W_{\rm E} &= {\cal W}_{\omega}.
\end{align}
Therefore, the dual theory is free from the double-counting problem by construction, and the local impurity contribution is excluded from the diagrams on the level of the bare propagators~\eqref{eq:barefermionpropagator}-\eqref{eq:barebosonpropagator}.
The DB relations up to this point are exact and derived without any approximations.

It is worth mentioning, that the non-interacting dual theory ($\tilde{V}=0$) is equivalent to EDMFT. However, even in the weakly-interacting limit of the original model, $U\to~0$, the fermion-boson vertex $\gamma^{2,1}$ is non-zero and equal to unity, as shown in Appendix~\ref{ap:dualtr} and previous works on the DB approach. Thus, the Dual Boson formalism explicitly shows that corrections to EDMFT are not negligible. Therefore, the dynamical mean-field level is insufficient for describing non-local bosonic excitations, because the interactions between the non-local fermionic and bosonic degrees of freedom are always relevant.

\subsection{Dual diagrams for the self-energy and polarization operator}
\begin{figure}[t!]
\includegraphics[width=0.95\linewidth]{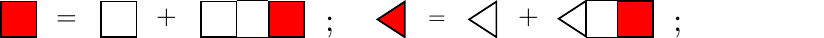}\\
\vspace{0.4cm}
\includegraphics[width=0.95\linewidth]{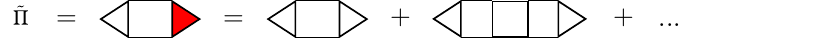}\\
\vspace{0.2cm}
\includegraphics[width=0.95\linewidth]{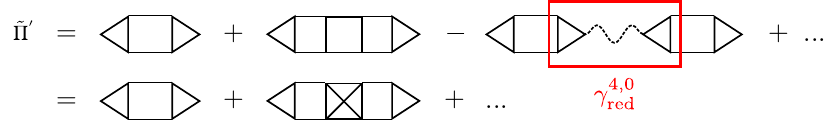}
\caption{Structure of the vertex corrections in different theories in case of one (top line) and two (bottom line) types of propagators. Solid straight and wave lines correspond to the Green's function and renormalized interaction of one type and the dashed lines to those of the second type respectively.}
\label{fig:PiPi'}
\end{figure}

The impurity vertices $\gamma^{n,m}$ are computationally expensive to calculate for large $n$ and $m$.
Practical DB calculations are usually restricted to $\gamma^{4,0}$ and $\gamma^{2,1}$, since that is sufficient to satisfy conservation laws and since processes involving higher-order vertices can be suppressed with the appropriate self-consistency condition~\cite{PhysRevB.93.045107}.

As it was shown above, the dual theory can be rewritten in terms of lattice quantities (see Eqs.~\eqref{eq:dual_G-Sigma}-\eqref{eq:dual_X-Pi}), where the dual diagrams are constructed in terms of only one type of bare propagators, i.e. the non-local part of EDMFT Green's function and renormalized interaction given by Eqs.~\eqref{eq:barefermionpropagator}-\eqref{eq:barebosonpropagator}. Local parts of the bare EDMFT propagators, namely $g_{\nu}$ and ${\cal W}_{\omega}$, are of the second type and hidden in the full local vertex functions of the impurity problem, which serve as the bare interaction vertices in dual space. Then, with the same logic presented in Section~\ref{sec:LocVertCor}, the DB self-energy and polarization operator in the ladder approximation are given by
\begin{align}
&\includegraphics[width=0.5\linewidth]{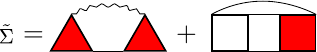}~,\label{eq:fulllS}\\
&\includegraphics[width=0.5\linewidth]{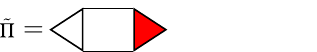}\hspace{-1.9cm},\label{eq:fulllP}
\end{align}
where the renormalized vertex functions are taken in the ladder approximation (see Fig.~\ref{fig:PiPi'} top line). Note, that here the splitting of propagators into the two parts is nominal and matters only for the dual theory when all diagrams are written in terms of only one non-local type of bare propagators. In general, the initial lattice theory works only with one type of Green's function and renormalized interaction, namely the bare EDMFT propagators, that for the local case we call as impurity $g_{\nu}$ or ${\cal W}_{\omega}$ and for non-local as dual $\tilde{G}_0$ or $\tilde{W}_0$. Since the dual theory gives the correction to the lattice quantities, the dual contributions $\bar{\Sigma}_{\kv\nu}$ and $\bar{\Pi}_{\qv\omega}$ introduced beyond EDMFT should be irreducible with respect both to the impurity and the dual propagators.

Let us turn to a more detailed explanation.
As was shown in Eqs.~\eqref{eq:dual_G-Sigma}-\eqref{eq:dual_X-Pi}, the lattice self-energy and polarization operator introduced beyond EDMFT are not given by the dual $\tilde{\Sigma}_{\kv\nu}$ and $\tilde{\Pi}_{\qv\omega}$ and have the form of Eqs.~\eqref{eq:Sigmap}-\eqref{eq:Pip}. Note, that the denominators in the expressions for $\bar{\Sigma}_{\kv\nu}$ and $\bar{\Pi}_{\qv\omega}$ have very important physical meaning. The DB theory works with the {\it full} vertex functions of impurity problem, that obviously contain one-particle reducible contributions. Therefore, the denominators in Eqs.~\eqref{eq:Sigmap}-\eqref{eq:Pip} exclude these one-particle reducible contributions from the diagrams for the self-energy and polarization operator in order to avoid the double-counting in the Dyson Eqs.~\eqref{eq:dual_G-Sigma}-\eqref{eq:dual_X-Pi}. Similar discussions were presented in Ref.~\cite{PhysRevB.88.115112} with regards to the DF approach.
\begin{figure}[t!]
\includegraphics[width=0.95\linewidth]{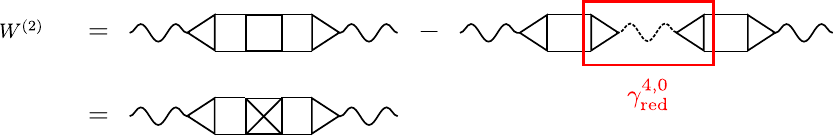}\\
\vspace{0.4cm}
\includegraphics[width=0.95\linewidth]{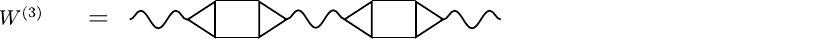}\\
\vspace{0.3cm}
\includegraphics[width=0.95\linewidth]{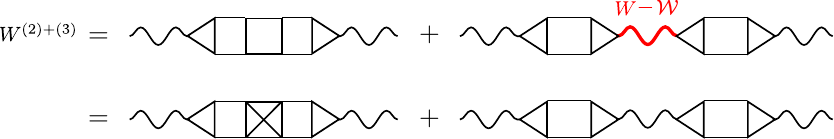}
\caption{Diagrammatic representation of the second and the third order contribution to the renormalized interaction.}
\label{fig:W}
\end{figure}

To show this more explicitly, let us consider the following example. The dual polarization operator in the form of full two-particle ladder can be written in a matrix form as (see the second line of Fig.~\ref{fig:PiPi'} for the diagrammatic representation)
\begin{align}
\tilde{\Pi}_{\kv\omega} = \frac{\gamma{}\tilde{G}\tilde{G}\gamma}{1+\left[\gamma{}\right]^{-1}\gamma^{4,0}\tilde{G}\tilde{G}\gamma},
\end{align}
where $\gamma^{4,0}$ is the full local four-point vertex of impurity problem. Using these relations, equation \eqref{eq:dual_X-Pi} can be rewritten as (see the third line of Fig.~\ref{fig:PiPi'})
\begin{align}
\bar\Pi^{\phantom{1}}_{\qv\omega} &= \frac{\gamma{}\tilde{G}\tilde{G}\gamma{}}{1+\left[\gamma{}\right]^{-1}\left(\gamma^{4,0}+\gamma{}{\cal W}_{\omega}\gamma\right)\tilde{G}\tilde{G}\gamma}.
\end{align}
Here
\begin{align}
&\gamma^{4,0}_{\rm irr}=\gamma^{4,0}+\gamma{\cal W}_{\omega}\gamma,\notag\\
&\includegraphics[width=3.1cm]{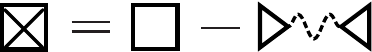}
\label{eq:red-irr-vertex}
\end{align}
is identically the irreducible part $\gamma^{4,0}_{\rm red}$ of full four-fermionic vertex function of impurity problem with respect to the renormalized interaction ${\cal W}_{\omega}$. Then the polarization operator $\bar{\Pi}$ introduced beyond EDMFT is nothing more than the normal dual polarization operator $\tilde{\Pi}$ taken in the form of the full dual ladder, but with irreducible four-point vertices $\gamma^{4,0}_{\rm irr}$ instead of the full vertices $\gamma^{4,0}$ of the impurity problem. Therefore, the exact relation \eqref{eq:Pip} automatically corrects the structure of the polarization operator, which is irreducible with respect to the dual renormalized interaction, to be also irreducible with respect to the impurity interaction ${\cal W}_{\omega}$.

Let us then compare the second and the third order term of diagrammatic expansion of Eq.~\eqref{eq:dual_X-Pi} shown in Fig.~\ref{fig:W},
\begin{align}
W^{(2)}_{\qv\omega} &= W^{\rm E}_{\qv\omega}\bar{\Pi}^{\phantom{1}}_{\qv\omega}W^{\rm E}_{\qv\omega}, \label{eq:W2}\\
W^{(3)}_{\qv\omega} &= W^{\rm E}_{\qv\omega}\bar{\Pi}^{\phantom{1}}_{\qv\omega}W^{\rm E}_{\qv\omega}\bar{\Pi}^{\phantom{1}}_{\qv\omega}W^{\rm E}_{\qv\omega}. \label{eq:W3}
\end{align}
After the substitution of the the second term of $\bar{\Pi}$ to Eq.~\eqref{eq:W2} and of the first term of $\bar{\Pi}$ to Eq.~\eqref{eq:W3} we get
\begin{align}
W^{(2)}_{\qv\omega}=&-W_{\rm E}\gamma{}GG\gamma^{4,0}_{\rm irr}GG\gamma{}W_{\rm E}, \label{eq:W2n}\\
W^{(3)}_{\qv\omega}=&\,\,\,\,\,\,\,W_{\rm E}\gamma{}GG\gamma{}{\cal W}_{\omega}\gamma{}GG\gamma{}W_{\rm E} \label{eq:W3n}\\
&+W_{\rm E} \gamma{}GG\gamma{}(W_{\rm E}-{\cal W}_{\omega})\gamma{}GG\gamma{}W^{\rm E}_{\qv\omega}, \notag\\
W^{(2)+(3)}_{\qv\omega} =&-W_{\rm E}\gamma{}GG\gamma^{4,0}_{\rm irr}GG\gamma{} W_{\rm E} \label{eq:W23n}\\
&+W_{\rm E} \gamma{}GG\gamma{}W_{\rm E}\gamma{}GG\gamma{}W^{\rm E}_{\qv\omega}.\notag
\end{align}
Then one can see, that the first term in Eq.~\eqref{eq:W3n} exactly gives the reducible contribution to the full four-point vertex function that was excluded from Eq.~\eqref{eq:W2n} by the denominator of $\bar{\Pi}$. If one neglects this denominator, it will immediately lead to the double-counting in Dyson Eq.~\eqref{eq:W23n}.

The same holds for the self-energy, where all contributions, coming from the denominator, give corrections to the six-point vertices $\gamma^{6,0}$ and $\gamma^{2,2}$ and remove the reducible contributions with respect to the local impurity Green's function $g_{\nu}$. Previous DB studies did not account for the six- and higher-point vertices, because they are negligibly small in both the large and small $U$ limits \cite{PhysRevB.93.045107}. Therefore, from one point of view, if the ladder approximation for the dual self-energy does not contain these six-point vertices, then the denominator in Eq.~\eqref{eq:Sigmap} should be neglected, because otherwise it will cancel the reducible terms in Dyson Eq.~\eqref{eq:dual_G-Sigma} with respect to the impurity $g_{\nu}$. On the other hand, one of the advantages of the DB formalism is the fact that all dual diagrams are written in terms of full impurity vertices instead of irreducible ones. Therefore, in the strong interaction limit, where the formal diagrammatic expansion cannot be performed, the full high-order vertices are small, which is not the case for the irreducible ones. Thus, writing the dual diagrams in terms of full vertices, it allows us to exclude the terms with the six-point vertices from the self-energy. Then, the presence of denominator in Eq.~\eqref{eq:Sigmap} helps to include irreducible contributions of the high-order vertices when their full contributions are negligibly small.

\subsection{DB$-$GW approach}

With the four-fermion vertex $\gamma^{4,0}$, the Dual Boson approach can obviously include more correlation effects than EDMFT$+$GW, at a significant computational cost.
However, it is also possible to construct a EDMFT++ approach from DB that does not require the full two-particle vertex. Taking $\gamma^{4,0}=0$, the fermion-boson vertex $\gamma_{\nu\omega}$ can be approximated as unity, as was discussed above, and the expressions for the dual $\tilde{\Sigma}_{\kv\nu}$ and $\tilde{\Pi}_{\qv\omega}$ operator are
\begin{align}
\tilde{\Sigma}^{\text{DB$-$GW}}_{\kv\nu}
&= -\sum\limits_{\qv\omega}\tilde{G}_{\kv+\qv,\nu+\omega}\tilde{W}_{\qv\omega},\\
\tilde{\Pi}^{\text{DB$-$GW}}_{\qv\omega} &= \,2\, \sum\limits_{\kv\nu}\tilde{G}_{\kv+\qv,\nu+\omega}\tilde{G}_{\kv\nu}.
\label{eq:dual_GW_limit}
\end{align}
We call this the DB$-$GW approximation. According to the above discussions, in this simplest case the denominator in Eqs.~\eqref{eq:Sigmap}-\eqref{eq:Pip} should be excluded, since we are interested in the contribution of only lower-order vertex functions, so we should take
\begin{align}
\bar{\Sigma}^{\phantom{1}}_{\kv\nu}&=\tilde{\Sigma}^{\rm DB-GW}_{\kv\nu},\\ \bar{\Pi}^{\phantom{1}}_{\qv\omega}&=\tilde{\Pi}^{\rm DB-GW}_{\qv\omega},
\end{align}
without the denominators presented in Eqs.~\eqref{eq:Sigmap}-\eqref{eq:Pip}.
Thus we see, that the EDMFT$+$GW and DB$-$GW approaches start with the same form of the self-energy and polarization operator diagrams and with similar propagators based on the same EDMFT quantities $G_{\rm E}$ and $W_{\rm E}$. The difference between the two approaches lies in the way double-counting is excluded from these diagrams, which for DB$-$GW case is shown in Eqs.~\eqref{eq:barefermionpropagator}-\eqref{eq:barebosonpropagator}. This results in different self-energies $\tilde\Sigma_{\kv\nu}$, and polarization operators $\tilde\Pi_{\qv\omega}$ that are used to treat non-local effects beyond the EDMFT in these two different cases. Since the local and non-local correlation effects are intertwined in a complicated way, it is more efficient to exclude double-counting already on the level of bare EDMFT Green's function and bare interaction in the dual formalism, rather than to remove the local contribution of the full diagram.
This happens naturally in the exact dual Hubbard--Stratonovich transformation.

It is worth mentioning, that the dual renormalized interaction $\tilde{W}_{\qv\omega}$ does not depend on the form of decoupling. As it is shown in Eq.~\eqref{eq:tildeW}, both $UV$-- and $V$-- decoupling forms lead to the same result $U_{\qv}-{\cal U}_{\omega}=~V_{\qv}-~\Lambda_{\omega}$ for the interaction accounted beyond the dynamical mean-field level in the DB theory. It is then easy to see, that the DMFT+GW theory in a $V$-- decoupling form excludes the impurity interaction in a proper way, since the dual renormalized interaction~\eqref{eq:tildeW} in the case $\Lambda_{\omega}=0$ has exactly the form of $V$-- decoupling. Due to the problems arising in the EDMFT+GW approach in the $UV$-- decoupling form mentioned in Appendix~\ref{app:GW} we take the renormalized interaction for the EDMFT$+$GW($\gamma$) theories in the form of $V$-- decoupling for the later comparison with DB results.
\begin{figure}[t!]
\includegraphics[width=0.9\linewidth]{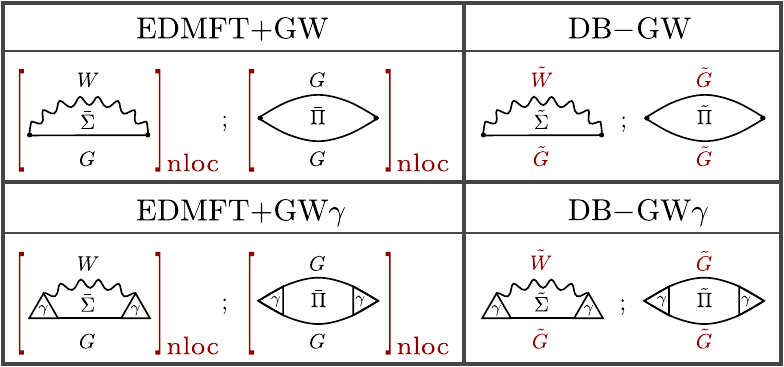}
\caption{Self-energy and polarization operator for  EDMFT++  approaches. The square brackets $[\ldots]_{{\rm nloc}}$ denote exclusion of the local part. DMFT$+$GW is not listed here, it has the same diagrams as EDMFT$+$GW and  only differs in their choice of ${\cal U}_\omega$.}
\label{fig:4methods}
\end{figure}

\subsection{Local vertex corrections in DB method}
To add vertex corrections to the DB$-$GW approach, one can take the second order diagrams for the dual self-energy $\tilde{\Sigma}^{{\rm GW}\gamma}=\bar{\Sigma}^{(2)}_{\kv\nu}$~\eqref{eq:Sigma2} and polarization operator $\tilde{\Pi}^{{\rm GW}\gamma}=\bar{\Pi}^{(2)}_{\qv\omega}$~\eqref{eq:Pi2}, which are dressed with the full local impurity fermion-boson vertices $\gamma_{\nu\omega}$ as
\begin{align}
\tilde{\Sigma}^{{\rm GW}\gamma}_{\kv\nu} &= -\sum\limits_{\qv\omega}\gamma^{\phantom{1}}_{\nu\omega}\tilde{G}_{\kv+\qv,\nu+\omega}\tilde{W}_{\qv\omega}\gamma^{\phantom{1}}_{\nu+\omega,-\omega}, \\
\tilde{\Pi}^{{\rm GW}\gamma}_{\qv\omega} &= \,2\, \sum\limits_{\kv\nu}\gamma^{\phantom{1}}_{\nu\omega}\tilde{G}_{\kv+\qv,\nu+\omega}\tilde{G}_{\kv\nu}\gamma^{\phantom{1}}_{\nu+\omega,-\omega}.
\label{eq:dual_GWd_limit}
\end{align}
Similarly to the DB$-$GW approach we neglect the denominator in Eqs.~\eqref{eq:Sigmap}-\eqref{eq:Pip} and repeat all calculations in the same way.

\begin{figure}[t!]
\includegraphics[width=0.9\linewidth]{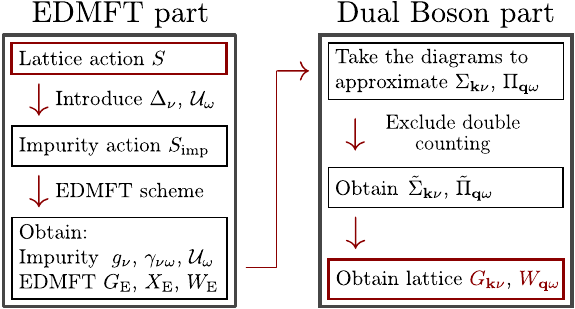}
\caption{The recipe to construct an EDMFT++ theory. DMFT$+$GW is obtained by taking ${\cal U}_\omega = U$ instead of determining it self-consistently.
}
\label{fig:recipe}
\end{figure}

The four approaches are summarized in Fig.~\ref{fig:4methods}, showing the self-energy and polarization operator diagram, where square brackets $[\ldots]_{{\rm nloc}}$ denote the exclusion of the local part.
The computational recipes for the all the EDMFT++ theories is shown in Fig.~\ref{fig:recipe}.

\section{Numerical results}

\begin{figure}[t!]
\includegraphics[width=0.9\linewidth]{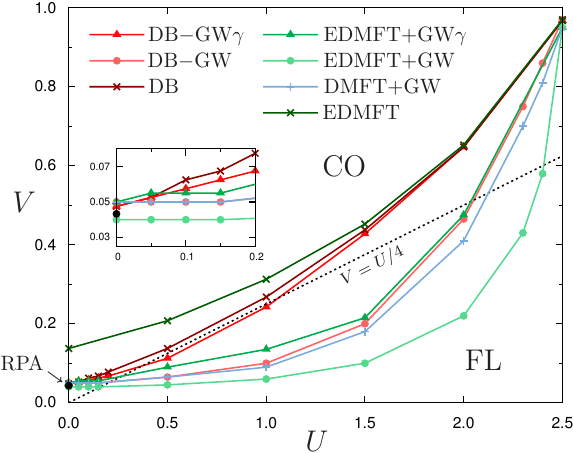}
\caption{$U-V$ phase diagram in EDMFT, DB and EDMFT++ theories at inverse temperature $\beta=50$. The dashed line shows $V=U/4$,
the dot at $U=0$ shows the starting point of RPA data. CO and FL denote charge-ordered and Fermi-liquid metallic phases, respectively. The EDMFT and DB data are taken from \cite{PhysRevB.90.235135}, EDMFT$+$GW data practically
coincides with results shown in \cite{PhysRevLett.109.226401, PhysRevB.87.125149} papers.}
\label{fig:phase}
\end{figure}

To test the EDMFT++ schemes, we study the charge-order transition in the square lattice Hubbard model, a popular testing ground for extensions of EDMFT~\cite{PhysRevB.87.125149,PhysRevB.90.195114,PhysRevB.90.235135}.
Here we show calculations where first $\Delta_\nu$ and $\Lambda_\omega$ are determined self-consistently on the EMDFT level for all schemes. Then, the non-local correlation effects are included. Having the same impurity problem as the starting point for all approaches allows us to compare clearly the the effect from the extensions only. We use $t=1/4$, $\beta=50$ and a $32\times{}32$ lattice.
The resulting phase boundary between the charge-ordered phase (CO) and the Fermi liquid (FL), determined in the same way as in Ref.~\cite{PhysRevB.90.235135}, is shown in Fig.~\ref{fig:phase}.
The checkerboard CO phase is characterized by a divergent charge susceptibility at the wave vector $\qv = (\pi, \pi)$. The phase boundary may therefore be located by looking for zeros of the inversed susceptibility $X^{-1}_{\omega=0, \qv=(\pi, \pi)}$. Note that the renormalized interaction $W_{\qv\omega}$ in DMFT$+$GW, EDMFT$+$GW and EDMFT$+$GW$\gamma$ approaches is taken in the form of the $V$-- decoupling as discussed above.

Since ordering is unfavorable for the interaction energy for $V<U/4$, the true phase boundary is expected to be above the $V=U/4$ line. Indeed, the full DB result is above this line~\cite{PhysRevB.90.235135}.
In all other EDMFT++ approximations with fewer correlation effects, the phase transition occurs at smaller $V$. The DB$-$GW$\gamma$ approximation performs best in this respect, and is close to the DB phase boundary for all values of $U$. The two approximations that include local vertex corrections via $\gamma_{\nu\omega}$ perform better than their counterparts without, and both DB based approaches outperform their EDMFT$+$GW counterpart.

At $U=0$, we expect the Random Phase Approximation (RPA) to give a reasonable prediction for the phase boundary. The RPA limit is recovered by all shown EDMFT++ approaches, but already 
at relatively small $U=0.5$, strong differences between the methods become clear.

In the opposite limit of large $U$, EDMFT itself starts to give an accurate phase boundary, since it accounts for all local effects and those are most important at large $U$. Both DB-based approaches converge to EDMFT at $U=2.5$, whereas both EDMFT$+$GW approaches give a phase boundary at the same, slightly smaller $V$.

We even observe that DMFT$+$GW performs better than EDMFT$+$GW, although it is simpler. Although DMFT$+$GW contains fewer correlation effects than EDMFT$+$GW, it is free from double-counting by construction. This clearly shows the huge role that double-counting can play. On the other hand, comparison of DMFT$+$GW and DB$-$GW schemes confirms the fact, that inclusion of bosonic correlations already on the impurity level is also very important and provides the better starting point for extending dynamical mean-field theory.

\begin{figure}[t!]
\centering
$\begin{array}{cc}
\includegraphics[trim = 0mm 0mm 0mm 0mm, clip, width=4.1cm]{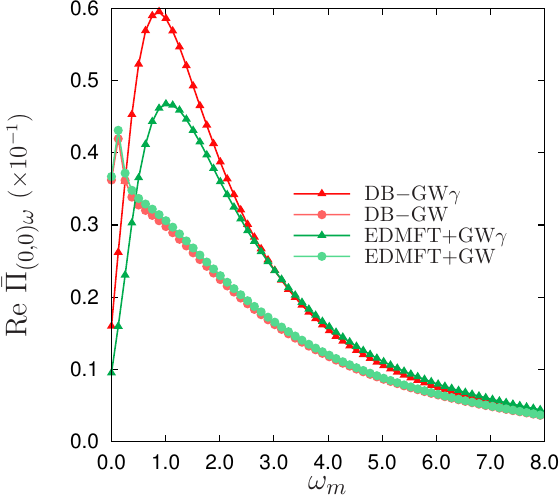}&
\includegraphics[trim = 0mm 0mm 0mm 0mm, clip, width=4.2cm]{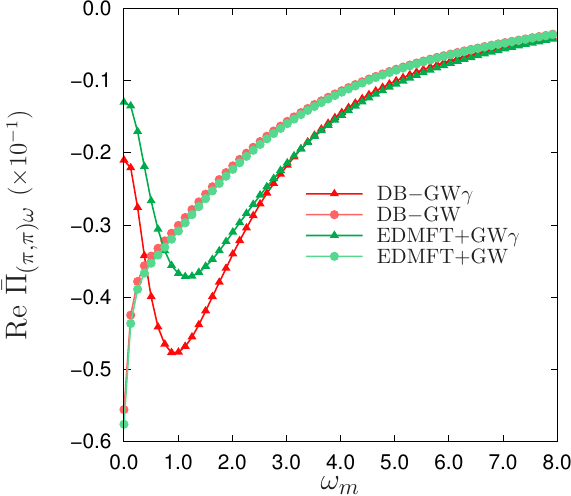}
\end{array}$
\caption{Frequency dependence of non-local Re $\bar{\Pi}_{\qv\omega}$ for momentum $\mathbf{k}=\left(0,0\right)$, $\mathbf{k}=\left(\pi,\pi\right)$ for on-site interaction $U=2.3$ and the nearest-neighbour interaction $V=0.2$.}
\label{SigmaNL}
\end{figure}

In Fig.~\ref{SigmaNL}, we show the polarization operator corrections $\bar{\Pi}_{\qv\omega}$ at high-symmetry $\qv$-points, according to the EDMFT$+$GW($\gamma$) and DB$-$GW($\gamma$) approaches. The results of the two approaches DB$-$GW and EDMFT$+$GW, that do not take into account the frequency dependent vertex function $\gamma$, are similar. The presence of the full local three-point vertex function in the diagrams significantly changes the results~\cite{PhysRevB.90.235135}. Moreover, the inclusion of the vertex function results in the different behavior of the polarization operator of the DB$-$GW$\gamma$ and EDMFT$+$GW$\gamma$ approaches. The dual contribution to the polarization operator in this case is larger. Therefore, using the dual way one excludes less contributions from the diagrams, than in the case of the EDMFT$++$ theories. Thus, the main difference in the approaches lies in their description of the collective excitations and comes from the different ways of treating the double-counting problem. 

The fermion-boson vertex exhibits less structure as the metallicity of the system is increased and becomes mostly flat as the phase boundary to the CO phase is approached~\cite{PhysRevB.90.235135}. The influence of non-local interaction $V$ on the three-point vertex function $\gamma_{\nu\omega}$ is shown in Fig.~\ref{fig:vertex}. The effects of the three-leg vertex are also visible in the non-local part of polarization operator in the difference between DB$-$GW and DB$-$GW$\gamma$ (or between EDMFT$+$GW and EDMFT$+$GW$\gamma$) approaches (see Fig.~\ref{SigmaNL}).

\section{Conclusions}
We have presented a recipe to create approximations beyond EDMFT that take into account non-local correlation effects while simultaneously avoiding double-counting issues.
By properly including non-locality we see an improvement in the phase boundary between the charge-ordered phase and the Fermi liquid.
Even in weakly and moderately interacting systems, the phase boundary is shifted significantly upwards compared to traditional EDMFT$+$GW.
In fact, EDMFT+GW is even improved upon by DMFT$+$GW, which neglects the effect of non-local interactions on the impurity model but does avoid double-containing.
This allows us to study the physics in a larger part of parameter space, where EDMFT$+$GW has undergone a spurious transition.
This is important for accurately determining the charge-ordering transition in real materials and in surface systems.

The approaches presented here work without requiring the computationally expensive full two-particle vertex. The frequency dependence of the much simpler fermion-boson vertex already contains most of the relevant physics, and including it via DB$-$GW$\gamma$ gives a phase boundary close to the full DB result. Without access to the fermion-boson vertex, deviations are bigger. In both cases, however, the dual way of treating the double-counting problem greatly improves the results.

The ladder dual boson approach can be derived from the dual functional, that automatically solve the complicated issue of the conservation laws~\cite{PhysRevB.93.045107}. For the future, it would be useful to obtain a similar functional description for the approximated theories presented in this work. 

\acknowledgments
The authors would like to thank Alexey Rubtsov for fruitful discussions and Lewin Boehnke and Andrey Katanin for useful comments.
E.A.S., E.G.C.P.v.L. and M.I.K. acknowledge support from ERC Advanced Grant 338957 FEMTO/NANO and from NWO via Spinoza Prize, A.I.L. from the DFG Research Unit FOR 1346, A.H. from the DFG via SPP 1459  and  computer  time  at  the  North-German  Supercomputing  Alliance  (HLRN).

\begin{appendix}

\section{Dual transformations}
\label{ap:dualtr}

\begin{figure}[t!]
\centering
$\begin{array}{cc}
\includegraphics[trim = 0mm 0mm 0mm 0mm, clip, width=4.1cm]{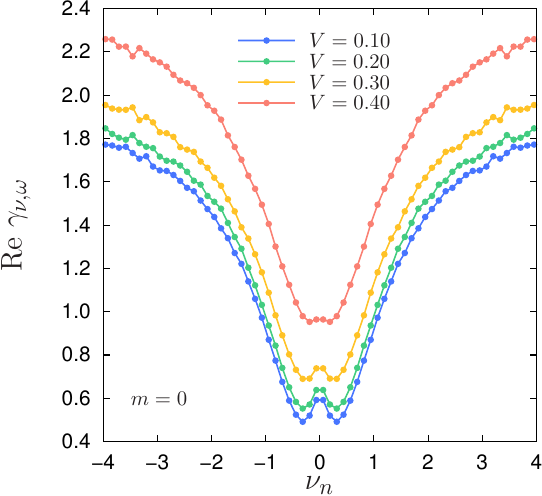}&
\includegraphics[trim = 0mm 0mm 0mm 0mm, clip, width=4.1cm]{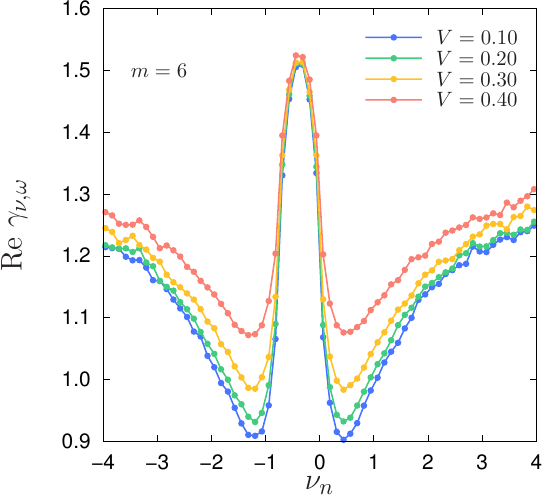}\\
\includegraphics[trim = 0mm 0mm 0mm 0mm, clip, width=4.1cm]{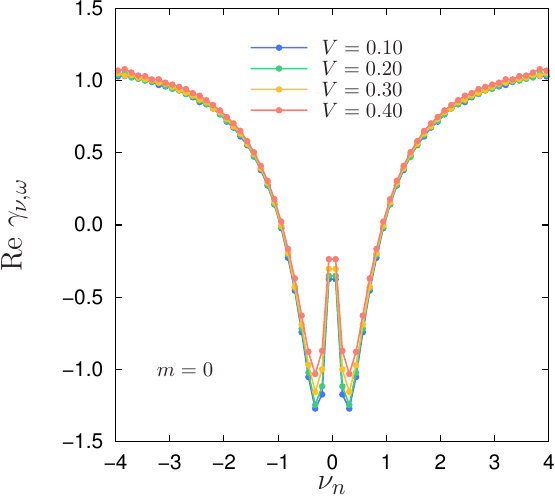}&
\includegraphics[trim = 0mm 0mm 0mm 0mm, clip, width=4.1cm]{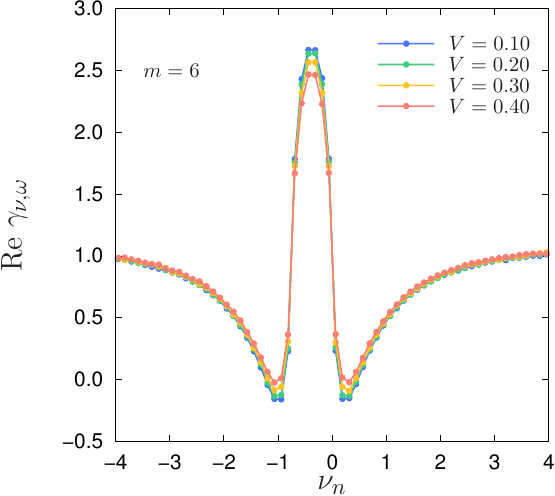}
\end{array}$
\caption{Local three-point vertex function $\gamma_{\nu\omega}$ for two bosonic frequencies $\omega_{m}=2m\pi/\beta$ with $m=0$ and $m=6$ for different values of nearest-neighbour interaction $V$ and local interaction $U=1.5$ (top line), and $U=2.3$ (bottom line).}
\label{fig:vertex}
\end{figure}

The dual transformations of the non-local part of the action $S_{\rm rem}$ can be made in the same way as in previous works on DB approach. In order to define the three-point vertex in the TRILEX way, here we introduce a different rescaling of the dual bosonic fields.
The partition function of our problem is given by
\begin{align}
Z=\int D[c^{*},c] \, e^{-S}
\end{align}
where the action $S$ is given by \eqref{eq:actionlatt}. Performing the Hubbard--Stratonovich transformations one can introduce the new ${\it dual}$ variables $f^{*},f,\phi$
\begin{align}
&e^{\,\sum\limits_{{\bf k}\nu\sigma} c^{*}_{{\bf k}\nu\sigma}[\Delta_{\nu\sigma}-\varepsilon_{\bf k}]c^{\phantom{*}}_{{\bf k}\nu\sigma}} = D_{f}\times\notag\\
&\int D[f^{*},f]\,e^{-\sum\limits_{{\bf k}\nu\sigma}\left\{ f^{*}_{{\bf k}\nu\sigma}[\Delta_{\nu\sigma}-\varepsilon_{\bf k}]^{-1}f^{\phantom{*}}_{{\bf k}\nu\sigma} + c^{*}_{\nu\sigma}f^{\phantom{*}}_{\nu\sigma} + f^{*}_{\nu\sigma}c^{\phantom{*}}_{\nu\sigma}\right\}},\notag\\
&e^{\,\frac12\sum\limits_{{\bf q}\omega} \rho^{*}_{{\bf q}\omega}[\Lambda_{\omega}-V_{\bf q}]\rho^{\phantom{*}}_{{\bf q}\omega}} \,\,\,= D_{\,b}\times\notag\\
&\int D[\phi]\,e^{-\frac12\sum\limits_{{\bf q}\omega}\left\{ \phi^{*}_{{\bf q}\omega}[\Lambda_{\omega}-V_{\bf q}]^{-1}\phi^{\phantom{*}}_{{\bf q}\omega} + \rho^{*}_{\omega}\phi^{\phantom{*}}_{\omega} + \phi^{*}_{\omega}\rho^{\phantom{*}}_{\omega}\right\}}.
\end{align}
Terms $D_{f} = {\rm det}[\Delta_{\nu\sigma}-\varepsilon_{\bf k}]$ and $D^{-1}_{\,b} = \sqrt{{\rm det}[\Lambda_{\omega}-V_{\bf q}]}$ can be neglected, because they does not contribute to expectation values. Rescaling the fermionic fields $f_{{\bf k}\nu\sigma}$ as $f^{\phantom{1}}_{{\bf k}\nu\sigma}g^{-1}_{\nu\sigma}$, the bosonic fields $\phi_{\qv\omega}$ as $\phi_{\qv\omega}\alpha^{-1}_{\omega}$, where $\alpha_{\omega}=~(1+{\cal U}_{\omega}\chi_{\omega})$, and integrating out the original degrees of freedom $c^*$ and $c$ we arrive at the dual action
\begin{align}
\tilde{S} &=
- \sum_{\kv\nu}f^{*}_{\kv\nu} \tilde{G}_{0}^{-1} f^{\phantom{*}}_{\kv\nu} - \frac{1}{2}\sum_{\qv\omega} \phi^{*}_{\qv\omega}\tilde{W}_{0}^{-1}\phi^{\phantom{*}}_{\qv\omega}+ \tilde{V}.
\end{align}
with the bare dual propagators
\begin{align}
\tilde{G}_{0} &= [g^{-1}_{\nu}+\Delta_{\nu}-\varepsilon_{\bf k}]^{-1} - g_{\nu} = G_{\rm E} - g_{\nu} ,
\\
\tilde{W}_{0} &= \alpha_{\omega}\left[[U_{\bf q}-{\cal U}_{\omega}]^{-1} - \chi_{\omega}\right]^{-1}\alpha_{\omega}
= W_{\rm E}-{\cal W}_{\omega},
\end{align}
and the dual interaction term $\tilde{V}$. The explicit form of the dual interaction can be obtained by expanding the $c^{*}$ and $c$ dependent part of partition function in an infinite series and integrating out these degrees of freedom as follows
\begin{align}
&\int
e^{-\sum\limits_{\nu\omega}\left\{c^{*}_{\nu}g^{-1}_{\nu}f^{\phantom{*}}_{\nu} + f^{*}_{\nu}g^{-1}_{\nu}c^{\phantom{*}}_{\nu} + \rho^{*}_{\omega}\alpha_{\omega}^{-1}\phi^{\phantom{*}}_{\omega} + \phi^{*}_{\omega}\alpha_{\omega}^{-1}\rho^{\phantom{*}}_{\omega}\right\}}\notag\\
&\,\,\,\,\,\,\,\,e^{-S_{\rm imp}[c^*,c]}\,D[c^{*},c] =
f^{*}_{\nu_1}f^{\phantom{*}}_{\nu_2}\av{c^{\phantom{*}}_{\nu_1}c^{*}_{\nu_2}}
g^{-1}_{\nu_1}g^{-1}_{\nu_2} \notag\\
&+\frac12\phi^{*}_{\omega_1}\phi^{\phantom{*}}_{\omega_2}
\av{\rho^{\phantom{*}}_{\omega_1}\rho^{*}_{\omega_2}}\alpha_{\omega_1}^{-1}\alpha_{\omega_2}^{-1} \notag\\
&-f^{*}_{\nu_1}f^{\phantom{*}}_{\nu_2}\phi^{*}_{\omega_3}
\av{c^{\phantom{*}}_{\nu_1}c^{*}_{\nu_2}\rho^{\phantom{*}}_{\omega_3}}g^{-1}_{\nu_1}g^{-1}_{\nu_2}\alpha_{\omega_3}^{-1}\notag\\
&+\frac14\,f^{*}_{\nu_1}f^{*}_{\nu_2}f^{\phantom{*}}_{\nu_3}
f^{\phantom{*}}_{\nu_4}\av{c^{\phantom{*}}_{\nu_1}c^{\phantom{*}}_{\nu_2}
c^{*}_{\nu_3}c^{*}_{\nu_4}}g^{-1}_{\nu_1}g^{-1}_{\nu_2}
g^{-1}_{\nu_3}g^{-1}_{\nu_4} + \ldots \notag\\
&=-f^{*}_{\nu}g^{-1}_{\nu}f^{\phantom{*}}_{\nu} - \frac12\phi^{*}_{\omega}\alpha_{\omega}^{-1}\chi^{\phantom{*}}_{\omega}\alpha_{\omega}^{-1}\phi^{\phantom{*}}_{\omega}\notag\\ &-f^{*}_{\nu_1}f^{\phantom{*}}_{\nu_2}\phi^{*}_{\omega_3}
\av{c^{\phantom{*}}_{\nu_1}c^{*}_{\nu_2}\rho^{\phantom{*}}_{\omega_3}}g^{-1}_{\nu_1}g^{-1}_{\nu_2}\alpha_{\omega_3}^{-1}\notag\\
&+\frac14\,f^{*}_{\nu_1}f^{*}_{\nu_2}f^{\phantom{*}}_{\nu_3}
f^{\phantom{*}}_{\nu_4}\av{c^{\phantom{*}}_{\nu_1}c^{\phantom{*}}_{\nu_2}
c^{*}_{\nu_3}c^{*}_{\nu_4}}g^{-1}_{\nu_1}g^{-1}_{\nu_2}
g^{-1}_{\nu_3}g^{-1}_{\nu_4} + \ldots \notag\\
&=e^{-\left\{ f^{*}_{\nu}g^{-1}_{\nu}f^{\phantom{*}}_{\nu} + \frac12\phi^{*}_{\omega}\alpha_{\omega}^{-1}\chi^{\phantom{*}}_{\omega}\alpha_{\omega}^{-1}\phi^{\phantom{*}}_{\omega}  +\tilde{V}\right\}}.
\end{align}
So the dual interaction has the form of an infinite expansion off the full vertices of the local impurity problem
\begin{align}
\tilde{V} = &\,\,f^{*}_{\nu_1}f^{\phantom{*}}_{\nu_2}\phi^{*}_{\omega_3}
\av{c^{\phantom{*}}_{\nu_1}c^{*}_{\nu_2}\rho^{\phantom{*}}_{\omega_3}}g^{-1}_{\nu_1}g^{-1}_{\nu_2}\alpha_{\omega_3}^{-1}-\notag\\
&\frac14\,f^{*}_{\nu_1}f^{*}_{\nu_2}f^{\phantom{*}}_{\nu_3}
f^{\phantom{*}}_{\nu_4}g^{-1}_{\nu_1}g^{-1}_{\nu_2}
g^{-1}_{\nu_3}g^{-1}_{\nu_4}
\big\{\av{c^{\phantom{*}}_{\nu_1}c^{\phantom{*}}_{\nu_2}
c^{*}_{\nu_3}c^{*}_{\nu_4}}- \notag\\
& \av{c^{\phantom{*}}_{\nu_1}c^{*}_{\nu_4}}
\av{c^{\phantom{*}}_{\nu_2}c^{*}_{\nu_3}} + \av{c^{\phantom{*}}_{\nu_1}c^{*}_{\nu_3}}
\av{c^{\phantom{*}}_{\nu_2}c^{*}_{\nu_4}}   \big\}+\ldots.
\end{align}
Here we define the three- and four-point vertex functions as ($\gamma_{\nu\omega}$ is the shorthand notation for the $\gamma^{2,1}_{\nu\omega}$),
\begin{align}
\gamma^{\phantom{1}}_{\nu\omega}&=
g^{-1}_{\nu}g^{-1}_{\nu+\omega}\alpha^{-1}_{\omega}
\av{c^{\phantom{*}}_{\nu}c^{*}_{\nu+\omega}\rho^{\phantom{*}}_{\omega}},\\
\gamma^{4,0}_{\nu\nu'\omega} &= g^{-1}_{\nu}g^{-1}_{\nu'}g^{-1}_{\nu'-\omega}g^{-1}_{\nu+\omega} \Big[\av{c^{\phantom{*}}_{\nu}c^{\phantom{*}}_{\nu'}c^{*}_{\nu'-\omega}
c^{*}_{\nu+\omega}}- \notag\\
&\hspace{3.5cm}g_{\nu}g_{\nu'}(\delta_{\omega}-\delta_{\nu',\nu+\omega})\Big],
\end{align}
with the simple connection between them
\begin{align}
\gamma^{\phantom{1}}_{\nu\omega} = \alpha_{\omega}^{-1}\sum_{\nu'}\big[1 - \gamma^{4,0}_{\nu\nu'\omega}g_{\nu'}g_{\nu'-\omega}\big].
\label{eq:ConGamma}
\end{align}
In the weakly-interacting limit, namely $U\to0$, the renormalization factor $\alpha_{\omega}$ goes to unity and the four-point vertex $\gamma^{4,0}$ is zero, as detailed in previous works~\cite{Rubtsov20121320, PhysRevB.90.235135, PhysRevB.93.045107} on the DB approach. Then, the three-point vertex can be reduced to its bare form $\gamma_0=1$. Frequency dependence of the full local three-point vertex function $\gamma_{\nu\omega}$ and the influence of non-local interaction $V$ is shown in Fig.~\ref{fig:vertex}.

Then, the two first terms in $\tilde{V}$ are given by
\begin{align}
\tilde{V}
&=\gamma^{\phantom{1}}_{\nu\omega}\,f^{*}_{\nu}f^{\phantom{*}}_{\nu+\omega}\phi^{*}_{\omega}
 + \frac14\,\gamma^{4,0}_{\nu\nu'\omega}\,f^{*}_{\nu}f^{*}_{\nu'}f^{\phantom{*}}_{\nu+\omega}
f^{\phantom{*}}_{\nu'-\omega}.
\end{align}
The dual Green's function $\tilde{G}_{\kv\nu} =~ -\av{f^{\phantom{*}}_{\kv\nu}f^{*}_{\kv\nu}}$ and renormalized dual interaction $\tilde{W}_{\qv\omega} =~ -\av{\phi^{\phantom{*}}_{\qv\omega}\phi^{*}_{\qv\omega}}$, as well as dual self-energy $\tilde{\Sigma}_{\kv\nu}$ and polarization operator $\tilde{\Pi}_{\qv\omega}$, can be obtained diagrammatically~\cite{Rubtsov20121320, PhysRevB.90.235135, PhysRevB.93.045107}. These dual quantities have usual connection
\begin{align}
\tilde{G}_{\kv\nu}^{-1} &= \,\tilde{G}^{-1}_{0}\, - \tilde{\Sigma}^{\phantom{1}}_{\kv\nu},\\
\tilde{W}_{\qv\omega}^{-1} &= \tilde{W}^{-1}_{0} - \tilde{\Pi}^{\phantom{1}}_{\qv\omega}.
\end{align}

Finally, lattice Green's function $G_{\kv\nu}$ and susceptibility $X_{\qv\omega}$ can be expressed in terms of dual propagators via exact relations
\begin{align}
G_{\kv\nu} = \,&-[\varepsilon_{\bf k}-\Delta_{\nu}]^{-1} \notag\\
&+[\varepsilon_{\bf k}-\Delta_{\nu}]^{-1}\, g^{-1}_{\nu}\,\tilde{G}^{\phantom{1}}_{{\bf k}\nu}\,g^{-1}_{\nu}[\,\varepsilon_{\bf k}-\Delta_{\nu}]^{-1},\\
X_{\qv\omega} = \, &- [U_{\bf q}-{\cal U}_{\omega}]^{-1} \notag\\
&+[U_{\bf q}-{\cal U}_{\omega}]^{-1}\alpha_{\omega}^{-1}\tilde{W}^{\phantom{1}}_{{\bf q}\omega}\alpha_{\omega}^{-1}[U_{\bf q}-{\cal U}_{\omega}]^{-1}.
\end{align}
One can also rewrite the last relation and obtain the relation for the full dual renormalized interaction
\begin{align}
\alpha^{-1}_{\omega}\tilde{W}^{\phantom{1}}_{{\bf q}\omega}\alpha^{-1}_{\omega} &= [U_{\bf q}-{\cal U}_{\omega}] +
[U_{\bf q}-{\cal U}_{\omega}]X_{\qv\omega}[U_{\bf q}-{\cal U}_{\omega}],
\label{eq:tildeW}
\end{align}
to show that the dual propagator $\tilde{W}_{{\bf q}\omega}$ is evidently a renormalized interaction in the non-local part of the action, where the impurity interaction is excluded on the level of the bare interaction.
It is worth mentioning, that for the case of $\Lambda_{\omega}=0$, which corresponds to the DMFT theory as a basis, the renormalized interaction is exactly that of the usual $V$-- decoupling.

\section{Comparison of the different decoupling schemes with the DB approach}
\label{app:GW}
As a consequence of the exact dual transformations presented in Appendix~\ref{ap:dualtr}, the renormalized interaction introduced beyond the DMFT when the bosonic hybridization function $\Lambda_{\omega}$ is equal to zero (i.e. $U_{\omega}=U$) should be taken in the form of $V$-- decoupling \eqref{eq:tildeW}. Contrary to DMFT, the impurity model in the EDMFT approach contains non-zero bosonic retarded interaction $\Lambda_{\omega}$, thus the renormalized interaction in EDMFT++ theories has neither $UV$--, nor $V$-- decoupling form. In this case the bare non-local interaction $U_{\qv}-{\cal U}_{\omega}$ for small $\Lambda_{\omega}$ (i.e. $U_{\omega}\simeq{}U$) is closer to $V_{\qv}$ then to $U_{\qv}=U+V_{\qv}$, and therefore in this paper we take $W_{\qv}$ in the form of $V$-- decoupling for all EDMFT++ theories.

One more argument to avoid treating the renormalized interaction in the $UV$-- decoupling form is the fact, that in this case EDMFT+GW reproduces the results of GW approach in the region close to the phase boundary. Indeed, the self-energy and polarization operator for the GW approach are given by Eqs.~\eqref{eq:Sigma_Pi_GW}-\eqref{eq:Sigma_Pi_GW1} respectively. The EDMFT$+$GW approach uses only non-local parts of these diagrams beyond the dynamical mean-field solution. They can be written as follows
\begin{align}
\bar{\Sigma}^{\rm E+GW}_{\kv\nu} &= -\sum\limits_{\qv\omega}\bar{G}^{\rm E+GW}_{\kv+\qv,\nu+\omega}\bar{W}^{\rm E+GW}_{\qv\omega}, \label{eq:SigmaUVd}\\
\bar{\Pi}^{\rm E+GW}_{\qv\omega} &= \,2\,\sum\limits_{\kv\nu}\bar{G}^{\rm E+GW}_{\kv+\qv,\nu+\omega}\,\bar{G}^{\rm E+GW}_{\kv\nu},
\end{align}
where $\bar{G}^{\rm E+GW}_{\kv\nu} = G_{\kv\nu}^{\phantom{1}}-g_{\nu}$, $\bar{W}^{\rm E+GW}_{\qv\omega}=~W^{\phantom{1}}_{\qv\omega} - {\cal W}^{\phantom{1}}_{\omega}$ are non-local parts of the full lattice Green's function and renormalized interaction respectively. Then, the full self-energy and polarization operator of the lattice problem can be written as
\begin{align}
\Sigma_{\kv\nu}^{\phantom{1}} &= \Sigma_{\rm imp}^{\phantom{1}} + \bar{\Sigma}^{\rm E+GW}_{\kv\nu},\\
\Pi^{\phantom{1}}_{\qv\omega} &= \Pi_{\rm imp}^{\phantom{1}} + \bar{\Pi}^{\rm E+GW}_{\qv\omega},
\end{align}
where
\begin{align}
\Sigma_{\rm imp} &= -\sum\limits_{\omega}g_{\nu+\omega}{\cal W}_{\omega}\gamma^{\phantom{1}}_{\nu\omega},\\
\Pi_{\rm imp} &= \,2\,\sum\limits_{\nu}g_{\nu+\omega}\,\,g_{\nu}\,\gamma^{\phantom{1}}_{\nu\omega},\label{eq:HedinPiImp}
\end{align}
are the exact self-energy and polarization operator of impurity problem written in the Hedin form. Then, one can rewrite the full lattice self-energy and polarization operator as
\begin{align}
\Sigma_{\kv\nu} &= -\sum\limits_{\qv\omega}G_{\kv+\qv,\nu+\omega}W_{\qv\omega} -\sum\limits_{\omega}g_{\nu+\omega}{\cal W}_{\omega}\left(\gamma^{\phantom{1}}_{\nu\omega}-1\right) \notag\\
&= \, \Sigma^{\rm GW}_{\kv\nu}-\sum\limits_{\omega}g_{\nu+\omega}{\cal W}_{\omega}\left(\gamma^{\phantom{1}}_{\nu\omega}-1\right), \label{eq:gwSigma}\\
\Pi_{\qv\omega} &= \,2\,\sum\limits_{\kv\nu}G_{\kv+\qv,\nu+\omega}\,G_{\kv\nu} + \,2\,\sum\limits_{\nu}g_{\nu+\omega}g_{\nu}\left(\gamma^{\phantom{1}}_{\nu\omega}-1\right) \notag\\
&= \, \Pi^{\rm GW}_{\qv\omega} + \,2\,\sum\limits_{\nu}g_{\nu+\omega}g_{\nu}\left(\gamma^{\phantom{1}}_{\nu\omega}-1\right).
\label{eq:gwPi}
\end{align}

Therefore, in the region where the value of the three-point vertex $\gamma_{\nu\omega}$ is close to the value of the bare three-point vertex $\gamma_{0}=1$, the EDMFT$+$GW approach reproduces the result of the GW method. Thus, the contribution of the exactly solvable impurity model in this region is lost. It happens, because one cancels the very big local contribution from the GW diagrams in order to avoid the double-counting problem, and then this local contribution suppresses the contribution of the local impurity model. It turns out that the EDMFT$+$GW approach cancels too much from the diagrams introduced beyond the dynamical mean-field level, and treating of the double-counting problem can be done in a better way. 

To see this, one can compare the dual way of exclusion of the double-counting with the $UV$-- decoupling scheme. Since the inner self-consistency for the diagrams beyond the dynamic mean-field level is used, it is hard to compare the resulting diagrams of these two approaches. Nevertheless, let us consider the polarization operator in the first iteration, when only the bare EDMFT Green's functions enter the diagrams. Studying the first iteration is sufficient since the non-local self-energy $\tilde{\Sigma}_{\kv\nu}$ is small in our region of interest. Then, one can see, that polarization operator for EDMFT+GW and DB$-$GW has the same form
\begin{align}
\tilde{\Pi}^{0}_{\qv\omega} &= \,2\,\sum\limits_{\kv\nu}\tilde{G}_{0}\,\tilde{G}_{0},
\end{align}
where $\tilde{G}_{0} = G_{\rm E}-g_{\nu}$. Then, one can obtain for the difference between the renormalized interactions used in EDMFT$+$GW and DB$-$GW the following relation
\begin{align}
&\left[W_{\qv\omega} - {\cal W}_{\omega}\right] - \tilde{W}_{\qv\omega} 
= \frac{W_{\rm E}}{1-\tilde{\Pi}^{0}_{\qv\omega}W_{\rm E}} - \frac{\tilde{W}_{0}}{1-\tilde{\Pi}^{0}_{\qv\omega}\tilde{W}_{0}} - {\cal W}_{\omega} \notag\\
&= {\cal W}_{\omega}[1-\tilde{\Pi}^{0}_{\qv\omega}\tilde{W}_{0}]^{-1}[1-\tilde{\Pi}^{0}_{\qv\omega}W_{\rm E}]^{-1}-{\cal W}_{\omega} \notag\\
&= {\cal W}_{\omega}\tilde{\Pi}^{0}_{\qv\omega}\left[ \tilde{W}_{\qv\omega}+W_{\qv\omega}+\tilde{W}_{\qv\omega}\tilde{\Pi}^{0}_{\qv\omega}W_{\qv\omega}
 \right].\label{eq:UVdecoupling}
\end{align}
Therefore, the self-energy \eqref{eq:SigmaUVd} in the form of $UV$-- decoupling additionally to the non-local dual contribution accounts for some diagrams that have local renormalized interaction ${\cal W}_{\omega}$ in their structure. In the Dual Boson formalism all local propagators are gathered in the local vertex functions of the impurity problem, including the local renormalized interaction ${\cal W}_{\omega}$, which is a part of the local four-point vertex $\gamma^{4,0}$. For example, the first term in the right hand side of Eq.~\eqref{eq:UVdecoupling} gives the following contribution to the self-energy,
\begin{align}
\includegraphics[width=0.6\linewidth]{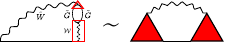}\,\,,
\label{eq:dc1}
\end{align}
which is a part of the dual diagram for the self-energy shown in Fig.~\ref{fig:Hedin}a).
The second term in the right hand side of Eq.~\eqref{eq:UVdecoupling}, when one takes only the local part of the EDMFT renormalized interaction in Eq.~\eqref{eq:nonlocal_Pi}, namely $W_{\qv\omega}=\frac{W_{\rm E}}{1-\tilde{\Pi}^{0}_{\qv\omega}W_{\rm E}}\sim \frac{{\cal W}_{\omega}}{1-\tilde{\Pi}^{0}_{\qv\omega}{\cal W}_{\omega}}$, is then equal to 
\begin{align}
\includegraphics[width=0.6\linewidth]{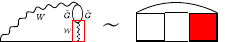},
\label{eq:dc2}
\end{align}
which is again a part of the dual diagram for the self-energy shown in Fig.~\ref{fig:Hedin}b). This fact leads to two important problems in the EDMFT$+$GW approach. First of all, these additional self-energy diagrams in case of $UV$-- decoupling presented above are very selective and account only for the local renormalized interaction ${\cal W}_{\omega}$ instead of the full local four-point vertex functions $\gamma^{4,0}$, as the DB approach does. This selective choice is not well-controlled and may result in over- or under-estimation of interaction effects. Also, the existence of the local propagators as a part of the non-local interaction shows that the EDMFT$+$GW approach in the form of $UV$-- decoupling is not able to separate local and non-local degrees of freedom in a proper way. This leads, in particular, to the double-counting problem in the next-order non-local diagrams introduced beyond EDMFT. Indeed, if one does not restrict himself to the simplest GW diagram accounted beyond the dynamical mean-field level and additionally includes the four-point vertex functions $\gamma^{4,0}$ in the diagrams for the self-energy (for example the diagrams shown in Fig.~\ref{fig:Hedin}b), then, as it was shown in Eqs.~\eqref{eq:dc1}-\eqref{eq:dc1}, the GW diagram \eqref{eq:SigmaUVd} for the self-energy would have contributions with the local ${\cal W}_{\omega}$, that would already be accounted for in these additional diagrams with the local four-point vertices.  

Let us study what happens in the region, where the impurity renormalized interaction ${\cal W}_{\omega}$ gives the main contribution in the full local four-point vertex $\gamma^{4,0}$. In this region the EDMFT$+$GW solution should be close to the Dual Boson ladder approximation with the self-energy and polarization operator diagrams shown in Fig.~\ref{fig:Hedin} a), b). Substituting ``$-{\cal W}_{\omega}$'' for the four-point vertex $\gamma^{4,0}$ in Eq.~\eqref{eq:ConGamma} and using Eq.~\eqref{eq:HedinPiImp} and the relation $\alpha^{-1} =~1-~\Pi_{\rm imp}\,{\cal U}_{\omega}$ one can get the trivial solution $\sum_{\nu}g_{\nu+\omega}g_{\nu}(\gamma_{\nu\omega}-1)=0$. Therefore, as it was shown in Eqs.~\eqref{eq:gwSigma}-\eqref{eq:gwPi}, in this region
EDMFT$+$GW in the $UV$-- decoupling form reproduces the result of the GW approach. 
In the other regions, where the bare vertex $\gamma_{0}=1$ does not give the main contribution to the full three-point vertex $\gamma_{\nu\omega}$, EDMFT$+$GW shows a result different from the GW approach, but unfortunately, it is not correct to approximate the full local vertex $\gamma^{4,0}$ by the local ${\cal W}_{\omega}$ there. 
As it was pointed out above, one of the advantages of the DB formalism is  that the full impurity vertices, in particular the full four-point vertex $\gamma^{4,0}$, are used in the dual diagrams for the self-energy and polarization operator. This full vertex $\gamma^{4,0}$ is small and consists of the two large contributions: reducible ($\gamma^{4,0}_{\rm red}$) and irreducible ($-\gamma{\cal W}_{\omega}\gamma$) with respect to renormalized interaction ${\cal W}_{\omega}$. These two contributions compensate each other as shown in Eq.~\ref{eq:red-irr-vertex}. If one accounts for only one large irreducible contribution to the vertex function, it leads to incorrect description of the collective excitations and problems mentioned above.  

Finally, one can rewrite Eq.~\eqref{eq:UVdecoupling} as follows
\begin{align}
\tilde{W}_{\qv\omega} 
= W_{\qv\omega} - {\cal W}_{\omega}\Big[1 + \tilde{\Pi}^{0}_{\qv\omega}\tilde{W}_{\qv\omega} + 
\tilde{\Pi}^{0}_{\qv\omega}&W_{\qv\omega} \notag\\ 
+\,\,\tilde{\Pi}^{0}_{\qv\omega}\tilde{W}_{\qv\omega}
\tilde{\Pi}^{0}_{\qv\omega}&W_{\qv\omega} \Big],
\end{align}
and see that DB excludes not the full local renormalized interaction ${\cal W}_{\omega}$ of the impurity model from the full lattice interaction $W_{\qv\omega}$, but the local interaction, that is renormalized by non-local polarization and non-local interactions $W_{\qv\omega}$ and $\tilde{W}_{\qv\omega}$. Therefore, the DB approach, which is free from the double-counting problem by construction, excludes less contributions from the full lattice renormalized interaction than the EDMFT$+$GW approach, and effects of the impurity model are not suppressed in our calculations.

\end{appendix}

\bibliography{DB_GW}

\end{document}